\documentclass[aps,preprint,showpacs]{revtex4}
\usepackage{amsfonts}
\usepackage{amsmath}
\usepackage{amssymb}
\usepackage{graphicx}
\usepackage{epsfig}
\begin{document}
\title{Back-to-back emission of the electrons in double photoionization of helium}
\author{M. Ya. Amusia $^{1,2}$, E. G. Drukarev $^{1,3}$, E. Z. Liverts $^1$}
\affiliation{$^1$The Racah Institute of Physics, The Hebrew
University of Jerusalem, 91904 Jerusalem, Israel\\ $^2$ A. F.Ioffe
Physico- Technical Institute, St. Petersburg 194021 Russia\\$^3$ B.
P. Konstantinov Petersburg Institute of Nuclear Physics, Gatchina,
St. Petersburg, 188300 Russia  }

\begin{abstract}
We calculate the double differential distributions and distributions in recoil momenta
 for the high energy non-relativistic double photoionization of helium. We show that the
 results of recent experiments is the pioneering experimental manifestation of the quasifree
 mechanism for the double photoionization, predicted long ago in our papers. This  mechanism
 provides a surplus in distribution over the recoil momenta at small values of the latter,
 corresponding to nearly "back-to-back" emission of the electrons. Also in agreement with
 previous analysis the surplus is due to the quadrupole terms of the photon-electron interaction.
 We present the characteristic angular distribution for the "back-to-back" electron emission.
 The confirmation of the quasifree mechanism opens a new area of exiting experiments,
 which are expected to increase our understanding of the electron dynamics and of the bound
 states structure. The results of this Letter along with the recent experiments
 open a new field for studies of two-electron ionization not
 only by photons but by other projectiles, e.g. by fast electrons or
 heavy ions.

\end{abstract}
\date{\today}
\maketitle

\section{ Introduction}
\label{S1}

In this Letter we calculate the distribution in recoil momenta for
the double photoionization of helium in the high energy
nonrelativistic limit. We calculate also the energy distribution in
the "back-to-back" configuration of the emitted electrons. Our
calculations are in agreement with the results of recent experiments
on the double ionization of helium by the photons with the energies
800 and 900 eV \cite{1}, \cite{2},  providing information on the
distribution in recoil momenta $q$ of the nucleus. Although the
groups, which carried out the experiments \cite{1}, \cite{2}, did
not present quantitative results, their experiments demonstrate that
the distribution of outgoing electrons obtains a surplus at small
$q$ of about 2 a.u. The kinematics of these experiments enables to
separate the non-dipole contributions at small values of $q$. Thus
the observed surplus is entirely due to the non-dipole terms. The
results of \cite{1}, \cite{2} prove the exitance of the
\textit{quasi-free mechanism} (QFM) of the double photoionization,
which was predicted many years ago \cite{3}.

By that time only two mechanisms of the process were known. In both
of them the electron, which interacted with the photon directly
obtained almost all the incoming photon energy $\omega$. In the
first, called \textit{shake-off} the secondary electron was pushed
to the continuum due to the sudden change of the effective field. In
the second, called \textit{knock-out} mechanism, the photoelectron
inelastically collides with the bound one, sharing the photon energy.
The two mechanisms could be clearly separated in the case of high
photon energies
\begin{equation}
\label{1}
\omega \gg I,
\end{equation}
with $I$ standing for the single-particle binding energy, when the
final state interactions between the outgoing electrons in the
shake-off mechanism can be neglected.

The key point of the third mechanism, predicted in \cite{3}, is that
the two electrons can absorb a photon almost without participation
of the nucleus. This is impossible in the shake-off and knock-out
mechanisms, since the single photoionization is not allowed for the
free electrons and thus in ionization, caused by a photon carrying
the energy $\omega$, momentum $q=(2\omega)^{1/2}$ (in atomic system
of units $\emph{e}=\hbar=m$, adopted in this paper) should be
transferred to the nucleus.

The QFM has several bright features. Before its prediction and
decades after the common wisdom was that the photoelectrons energy
spectrum curve has an ${\bf U}$ shape with high maxima at the edge
regions of the spectra. The QFM has predicted a local maximum at the
center of the energy distribution leading to ${\bf W}$ shape.
Another feature of QFM is that the contribution of it decreased with
energy slower than the contributions of the other mechanisms. Thus,
the account of QFM leads to the breakdown of the high energy
non-relativistic asymptotic of the double-to single photoionization
cross sections ratio. Also, the QFM requires going beyond the dipole
approximation, since there is no dipole moment of the two-electron
system at ${\bf q}=0$). One can see \cite{5} for more details.

Although the paper \cite{3} was cited rather often since its
publication,  QFM was for a long time not treated seriously by the
physical community as a two-electron photoionization mechanism. For
example, the QFM was not even mentioned in the review paper of
Dalgarno and Sadeghpour \cite{5a}. Attempts were undertaken to check
the QFM effects in purely computer calculations. These attempts fail
to confirm the existence of the QFM. Later it was understood
\cite{6} that the QFM is extremely sensitive to the analytical
properties of the initial state wave functions. In particular, it
cannot be reproduced in computations with uncorrelated electron
wavefunctions, which were used in the calculations, mentioned above.

Other developments were even more dramatic. Some of the calculations
lead to the ${\bf W}$ shape of the spectrum (see, \textit{e.g.}
\cite{5b}) even in the dipole approximation. It was shown, however,
in \cite{7} that the central peak there was spurious, being entirely
a consequence of oversimplified approximations for the wave
functions of either initial or the final states. The consistent
approach provided cancellation of spurious terms and restoration of
the ${\bf U}$ shape of the spectrum in the dipole approximation.

\section{The quasifree mechanism}
\label{S2}

If the condition (\ref{1}) is fulfilled, in the single
photoionization process the momentum $q$ exceeds strongly the
characteristic binding momentum $\eta$. However, in the double
photoionization there is a kinematical region, where the recoil
momentum $q$ can be as small as $\eta$. Following the general
analysis of Bethe \cite{4}, one can expect the increasing of the
differential cross section in this region. It happens because the
bound electrons are localized mainly near their Bohr orbits with the
radii $r_b \sim 1/\eta$. Each act of transferring larger momenta
requires going to the smaller distances to the nucleus, where the
electron density is smaller, leading to a smaller value of the
amplitude.

In the rest frame of the initial atom the recoil momentum is
\begin{equation}
\label{2}
{\bf q}={\bf k}-{\bf p_1}- {\bf p_2}.
\end{equation}
Here ${\bf k}$ is the photon momentum, ${\bf p_i}$ are the momenta
of the outgoing electrons.

Except the edge region of the spectra both $\varepsilon_i \gg I$, and thus $p_i \gg \eta$. Hence,
the QFM condition
\begin{equation}
\label{3}
\eta \leq q \ll p_{1,2},
\end{equation}
means that large momenta ${\bf p_i}$ almost compensate each other.
Hence, they are emitted mostly "back-to-back, with $t \equiv ({\bf
p_1} \cdot {\bf p_2})/p_1p_2$ close to $-1$. The amplitude is large
for $|t+1| \sim I/\omega \ll 1$. The condition (\ref{3}) can be
satisfied if the difference between the energies of outgoing
electrons $\varepsilon_i$ is small enough:
\begin{equation}
\label{4}
\beta \equiv \frac{|\varepsilon_{1}-\varepsilon_{2}|}{E}\leq
\sqrt\frac{2I}{E}; \quad E=\varepsilon_{1}+\varepsilon_{2},
\end{equation}
with $E=\varepsilon_{1}+\varepsilon_{2}$ the total energy carried by electrons.

As we have seen earlier, there is no dipole contribution in exactly
free kinematics with $q=0$. Such a process is caused by the
quadrupole and higher  multipole terms. In the quasifree kinematics
there is a non-varnishing dipole term proportional to $({\bf e}{\bf
q})$. However, it is strongly suppressed \cite{6}, and the
quadrupole terms do dominate for $\omega \geq $ 800 eV. Anyway, in
the experiment, described in \cite{1} they detect the recoiling ions
moving perpendicular to the polarization direction. This entirely
eliminates the contribution of the dipole terms.

In the QFM the two bound electrons exchange large momenta in the
initial state. Thus, they approach each other at small distances
$r_{12} < r_{b}$, while their distances from the nucleus is still of
the order of the Bohr orbit. Hence, it is reasonable to attribute
the QFM amplitude to the properties of the initial state wave
function $\psi(r_1,r_2, r_{12})$ at $r_{12}=0$. It was shown in
\cite{7} that the amplitude contains the factor $\partial
\psi/\partial r_{12}$ at $r_{12}=0$, which is connected to the
function
\begin{equation}
\label{5}
\phi(r)\equiv \psi(r,r,0)
\end{equation}
by the cusp condition \cite{8}.

\section{Distribution in recoil momenta}
\label{S3}
Now let us calculate the QFM amplitude of the high energy
nonrelativistic double photoionization.

Interactions of the outgoing electrons with the nucleus are determined
by their Sommerfeld parameters $\xi_i= Z/p_i$. Since both
$\varepsilon_i \gg I$, at the first step we can neglect interactions between the
outgoing electrons and the nucleus \cite{9}. Direct calculation
provides
\begin{equation}
\label{6}
\frac{d^2 \sigma}{dq^2d\varepsilon_1}=\frac{128}{15}\frac{\omega}{c E^3}S^2(q^2),
\end{equation}
with
\begin{equation}
\label{7}
S(q^2)=\int d^3r \phi(r)\exp{(-i({\bf q}\cdot {\bf r}))},
\end{equation}
with $\phi(r)$ defined by Eq.(5).

The analytical expressions, approximating very precise wave functions \cite{10} at $r_{12}=0$
were obtained in \cite{11}, \cite{12}. These functions work as well for approximating the
improved wave functions obtained in \cite{13}. In the simplest case \cite{11}
\begin{equation}
\label{8}
\phi(r)=\phi(0)\exp{(-2Z r)},
\end{equation}
with $Z$ being the charge of the nucleus. This provides
 \begin{equation}
 \label{9}
S(q^2)=\frac{16\pi Z\phi(0)}{(q^2+4Z^2)^2}.
\end{equation}
For the functions obtained in \cite{10,13} $\phi(0)\simeq 1.37$. Thus
indeed the distribution in recoil momentum $q$ has a surplus at
small $q \ll p_i$. To obtain the distribution $d \sigma/dq^2$, one
should integrate the distribution (\ref{6}) over $\varepsilon_1$, having
in mind that $q \geq |p_1-p_2|$.
In actual calculations, instead of (\ref{8}) we employ combination of two exponential
terms \cite{12} which gives a very accurate approximation of the exact wave function
at the electron-electron coalescence line.

In the experiments \cite{1,2} the parameters $\xi_i$ of the outgoing
electrons are of the order $1/3$. Thus, it is desirable to avoid
expansion in $\xi_i$, taking into account interaction with the
nucleus. In this case the factor $\exp{(-i({\bf q}\cdot {\bf r}))}$
in the integrand of Eq.(\ref{7}) should be replaced by the product
of the two continuum Coulomb functions. The integral can be
evaluated analytically by employing the technique, developed in
\cite{14}. Finally we obtain
\begin{equation}
\label{10}
\frac{d^2 \sigma}{dq^2d\varepsilon_1}=\frac{128}{15}\frac{\omega}{c E^3}S^2(q^2)F(\xi_i, q^2).
\end{equation}
The function $F$ with $F(\xi_1=0, \xi_2=0, q^2)=1$ has a simple analytical form.

These equations enable to obtain the angular distribution at the
point of exactly "back-to-back" emission by presenting
\begin{equation}
\label{11}
\frac{d^2 \sigma}{dtd\varepsilon_1}=2p_1p_2\frac{d^2 \sigma}{dq^2d\varepsilon_1}.
\end{equation}
We calculate the distributions $d^2 \sigma/dt d\varepsilon_1$ and $d
\sigma/dt$ at the point of exactly "back-to-back" emission $t=-1$.
In Fig.\ref{1} we provide example of the distribution $d^2\sigma/dt d\varepsilon_1$ for
the energy $\omega$=900 eV employed in
\cite{2}. One can see that the main contribution to $d \sigma/dt$
comes from $\beta \leq 0.3$ in agreement with Eq.(4). In Fig \ref{2}. we
show the dependence of the distribution $d \sigma/dt$ on the photon
energy in the region near 1 keV. At $\omega$=900 eV we find $d
\sigma/dt=0.52 barn$. Since the important interval of $t$ is $I/\omega
\approx 0.06$ the contribution to the total cross section is $0.03
b$ in agreement with \cite{15}.

 \section{Summary}
 \label{S4}

 We have calculated the distributions $d^2 \sigma/dq^2 d\varepsilon_1$ and $d \sigma/dq^2$ for the
 non-relativistic high energy double photoionization. Our results are consistent with those of
 the recent experiments \cite{1}, \cite{2}. Distribution in recoil momentum $q$ has a surplus
 at small $q$ caused by the quadrupole terms of electron-photon interaction. Thus, the existence
 of quasifree mechanism predicted long ago \cite{3} is confirmed. This
 opens a new area for experimental investigations of this mechanism.
 Note that the relative role of the QFM grows with the photon energy increase, and its
 manifestation for $\omega$ beyond the keV region is expected to be even more prominent.
 It is expected that the corresponding experimental and theoretical investigations will add much
 to our knowledge of the electron dynamics in the process of two-electron ionization and
 of the structure of the bound states wave functions.

 We dream that further research will move into relativistic region $\omega\geq c^{2}$
 thus disclosing the fine structure of the central peak of the energy distribution, caused by
 the non-dipole nature of the QFM. We hope also that contribution of
 the QFM to the total cross section, resulting in a slope of the double-to-single photoionization
 ratio will be measured. We expect the detailed investigation of the really relativistic case,
 where the QFM contribution should become as important as that of
 the shake-off and even much overcome it.

 Also, investigation of the QFM enables to clarify behavior of the wave function of the atom of
 helium near the singular electron-electron coalescence point. Besides the purely theoretical
 interest, this is important for precise computations of the atomic characteristics.
 Recall that the proper treatment of the three-particle coalescence point enabled to diminish
 strongly the number of parameters in the bound state wave functions.

 We expect that the results, presented in this Letter along with the recent experiments
 will stimulate studies of two-electron ionization
 by the other types of projectiles, such as the fast electrons or
 heavy ions, where along with quadrupole, monopole terms will contribute at least not less.

 The work was supported by the MNTI-RFBR grant 11-02-92484. One of us (EGD) thanks for
 hospitality during the visit to the Hebrew University of Jerusalem.

\begin{figure}
\begin{center}
\caption{Energy distribution for the "back-to-back" emission
($t=-1$) presented by Eq.(\ref{11}) for $\omega=900 eV$ considered
in \cite{2}. The value $d^2\sigma/d{\varepsilon_1}dt$ is given in
$barn / eV$.} \epsfxsize=15cm\epsfbox{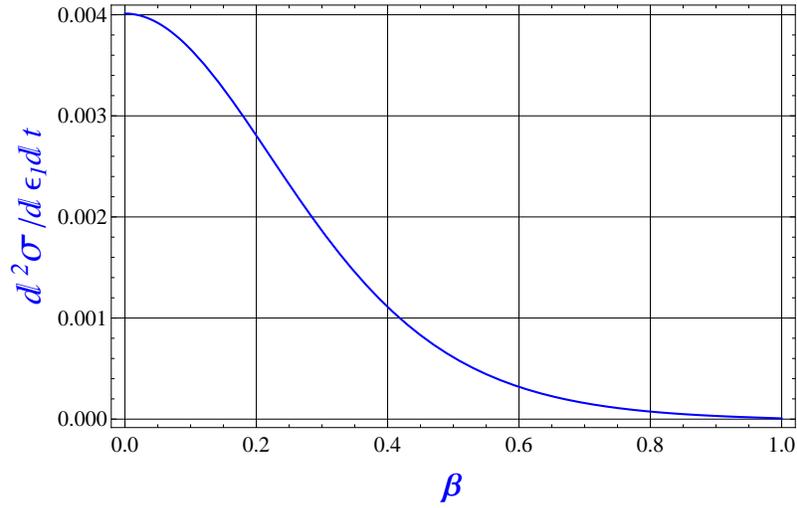} \label{F1}
\end{center}
\end{figure}

\begin{figure}
\begin{center}
\caption{Dependence of the differential distribution $d\sigma/dt$ at
$t=-1$ on the photon energy in keV region. The value of $d\sigma/dt$
is given in barns.} \epsfxsize=15cm\epsfbox{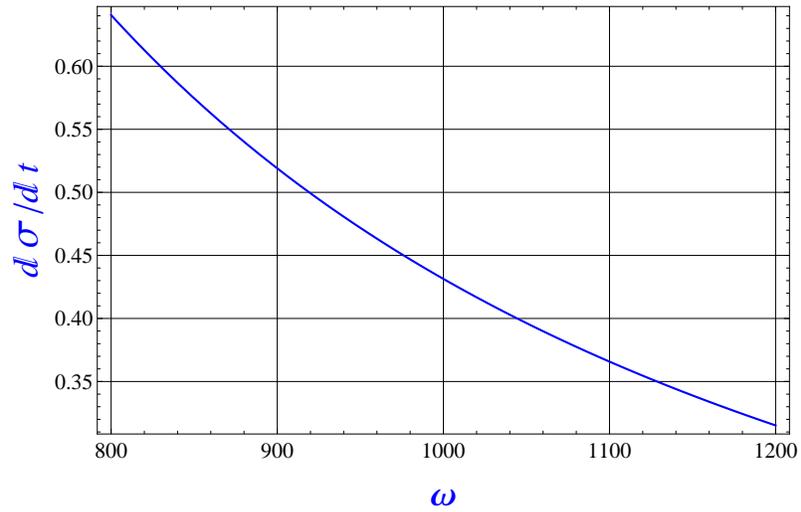} \label{F2}
\end{center}
\end{figure}

\end{document}